\documentclass[preprint]{ptptex}
\usepackage{graphicx,color}
\preprintnumber[3cm]{DPNU-03-28}
\title{Instability of Massive Scalar Fields in Kerr-Newman Spacetime}
\author{Hironobu \textsc{Furuhashi}\footnote{E-mail:\texttt{~hironobu@gravity.phys.nagoya-u.ac.jp}} and Yasusada \textsc{Nambu}\footnote{E-mail:\texttt{~nambu@gravity.phys.nagoya-u.ac.jp}}} 

\inst{Department of Physics, Graduate School of Science, Nagoya
University, Chikusa, Nagoya 464-8602, Japan}
\abst{
  We investigate the instability of charged massive scalar fields in
  Kerr-Newman spacetime. Due to the super-radiant effect of the
  background geometry, the bound state of the scalar field is
unstable, and its amplitude grows in time. By solving the
  Klein-Gordon equation of the scalar field as an eigenvalue problem,
  we numerically obtain the growth rate of the amplitude of the
  scalar field. Although the dependence of the scalar field mass and
  the scalar field charge on this growth rate agrees with the
  result of the analytic approximation, the maximum value of the
  growth rate is three times larger than that of the analytic
  approximation.  We also discuss the effect of the electric charge on
  the instability of the scalar field.
}

\date{\today}
\begin{document}
\maketitle
%%%%%%%%%%%%%%%%%%%%%%%%%%%%%%%%%%%%%%%%%%%%%%%
\section{INTRODUCTION}
%%%%%%%%%%%%%%%%%%%%%%%%%%%%%%%%%%%%%%%%%%%%%%%
The propagation of waves in black hole spacetime has been studied by
many researchers in the context of black hole
perturbations\,\cite{kokkotas99}. Super radiance is one type of
interesting phenomena related to this
topic\,\cite{zeldovich70,zeldovich71,zeldovich72,frolov}. Considering
the situation that a wave impinges on a black hole, the incident wave is
partially reflected by the potential barrier of the black hole, due to
the centrifugal force, and is scattered to infinity, while part of the
wave penetrates the potential barrier and is absorbed by the black hole.
Thus, one would expect that the amplitude of the scattered waves is always smaller than that of the incident wave.
However, this is not necessarily true in the Kerr-Newman geometry.
If the frequency of the incident wave satisfies the so-called
super-radiant condition, the reflected wave is amplified, and its
amplitude becomes larger than that of the incident wave; that is, the
reflection rate can be greater than unity.  Using this amplification
mechanism, it is possible to extract the rotation and the
electro-static energy of Kerr-Newman black holes. 

As an application of the super radiance, Press and Teukolsky proposed
the ``black hole bomb''\cite{press72}. They considered the situation in which
mirrors surround a Kerr black hole. Then, the scattered wave with
super-radiant amplification is reflected back to the black hole by the
mirrors, and the wave is amplified again. In this way, the amplitude of the
wave grows exponentially in time and becomes unstable. Damour {\it et
  al.}\,\cite{damour76} have shown that a black hole bomb can be
realized by using a charged massive scalar field in Kerr-Newman
spacetime. The effective potential of the massive scalar field has a
local minimum, and the scalar field wave can be confined in this potential
well. Hence, the mass of the scalar field plays the role of the mirrors,
which is necessary to cause the instability of the scalar field.
According to their analysis, the instability of the scalar field
is realized if the bound state and the super-radiant condition are both
satisfied.

To evaluate the growth rate of the scalar field, we must solve the
Klein-Gordon equation in the black hole geometry. By imposing ingoing
boundary conditions at the black hole horizon and regular boundary conditions
at infinity, the problem of obtaining the unstable mode reduces to an
eigenvalue problem.  For Kerr spacetime, the growth rate of the
unstable modes for a scalar field with a small mass $\mu\ll 1/M$
(where $M$ is the mass of the black hole) was obtained by using the asymptotic
matching method\,\cite{detweiler80}, and that for a large mass
$\mu\gg 1/M$ was obtained by using the WKB
approximation\,\cite{zouros79}.  However, these analyses do not cover
the parameter region in which $\mu M \sim 1$, where the growth rate is expected
to take its maximum value.

In this paper, we aim to obtain the growth rate of a charged
massive scalar field with mass satisfying $\mu M\lesssim 1$ in Kerr-Newman
spacetime by solving the eigenvalue problem numerically. The paper is
organized as follows.  In \S\ref{sec-analytic}, we estimate the
growth rate of a scalar field with $\mu M\ll 1$ and $q Q\ll 1$ in
Kerr-Newman spacetime using the method of Detweiler\cite{detweiler80}.
Then, in \S\ref{sec-numerical}, we introduce our numerical method
and obtain the growth rate caused by the instability and its parameter
dependence for $\mu M \lesssim 1$. Section~\ref{sec-summary} is
devoted to a summary and conclusion.

We use the units in which $G=c=\hbar=1$ throughout the paper.
 
%%%%%%%%%%%%%%%%%%%%%%%%%%%%%%%%%%%%%%%%%%%%%%%%%%%%%%%
\section{Analytic approach\label{sec-analytic}}
%%%%%%%%%%%%%%%%%%%%%%%%%%%%%%%%%%%%%%%%%%%%%%%%%%%%%%%
In this section, we calculate the growth rate of
the scalar field using the asymptotic matching method used by
Detweiler\,\cite{detweiler80}  in Kerr-Newman spacetime.

The Kerr-Newman metric in the Boyer-Lindquist coordinates is 
%%%
\begin{align}
  & ds^2=-\left(1-\frac{2Mr-Q^2}{\Sigma}\right)dt^2-
        \frac{(2Mr-Q^2)2a\sin^2\theta}{\Sigma}dtd\phi \notag\\
 &\qquad+\frac{\Sigma}{\Delta}dr^2+\Sigma d\theta^2
  +\left(r^2+a^2+\frac{(2Mr-Q^2)a^2\sin^2\theta}{\Sigma}\right)\sin^2\theta 
  d\phi^2,
    \label{eq:KN} \\
  & \Delta=r^2-2Mr+a^2+Q^2,\quad\Sigma=r^2+a^2\cos^2\theta,
  \notag
\end{align}
%%%
where $M$ is the mass, $a$ is the angular momentum and $Q$ is the
electric charge of the black hole. The locations of the horizons
$r_{\pm}$ are given by the roots of the equation $\Delta=0$. The
Klein-Gordon equation for a charged scalar field with mass $\mu$ is
%%%
\begin{align}
  & (\nabla^\alpha-iqA^\alpha)(\nabla_\alpha-iq A_\alpha)\Psi=\mu^2\Psi,
  \label{eq:KG} \\
  & A_\alpha=\left(-\frac{rQ}{\Sigma}, 0, 0,
  \frac{aQr}{\Sigma}\sin^2\theta\right), \notag
\end{align}
%%%
where $\nabla^\alpha$ is the covariant derivative in the Kerr-Newman
geometry, and $q$ is the charge of the scalar field. Equation
\eqref{eq:KG} is separable in terms of the spheroidal harmonics $S(\theta)$:
%%%
\begin{equation}
  \Psi=\psi(r)S(\theta)\exp(i(-\omega t+m\phi)).
\end{equation}
%%%
The radial function $\psi(r)$ satisfies the relation
%%%
\begin{align}
  &\Delta\frac{d}{dr}\Delta\frac{d\psi}{dr}
  +\left[-\Delta(\mu^2r^2+\lambda)+
  \left\{(r^2+a^2)\omega-ma-qQr\right\}^2\right]\psi=0,
  \label{eq:radial}
\end{align}
with
\begin{align}
 & \lambda=l(l+1)-2ma\omega+(a\omega)^2+O(a^2\mu^2(1-\omega^2/\mu^2)), \notag 
\end{align}
%%%%
where $l, m$ are integers and $|m|\leq l$. We assume $l \geq 1$. (For
$l=0$, there is no  centrifugal force, and the bound state of the
scalar field does not exist.) We are interested in the eigen-mode whose
frequency is nearly equal to the mass of the scalar field, i.e.
for which $\omega\sim\mu$. As in this case we have $|1-\omega^2/\mu^2|
\ll 1$, we can realize the relation $O(a^2 \mu^2(1-\omega^2/\mu^2)) \sim
0$, and then the separation constant is given by 
%%%
\begin{equation}
\label{eq:sepa-const}
 \lambda=l(l+1)-2ma\omega+(a\omega)^2.
\end{equation}
%%%
We solve Eq.\eqref{eq:radial} using regular boundary conditions at
infinity and ingoing boundary conditions at the black hole
horizon. To apply the asymptotic matching method, we need to assume that
the parameters satisfy the conditions 
%%%
\begin{equation}
O(|\omega M|)=O(\mu M)=O(q Q)\equiv O(\epsilon),\quad \epsilon \ll 1,
\end{equation}
%%%
and that they can thus be treated as small parameters. The angular momentum of
the black hole is assumed to satisfy $O(a/M)=1$.

For large $r$, i.e. $r\gg r_{+}$, Eq.\,\eqref{eq:radial} reduces to
\begin{equation}
\label{out-difeq}
  \frac{d^2}{dr^2}(\Delta^{1/2} \psi)
   +\left[\omega^2-\mu^2+\frac{2(2M\omega^{2}-M\mu^2-qQ\omega)}{r} 
    -\frac{l(l+1)+\epsilon ^{2}}{r^2}\right](\Delta^{1/2}\psi)=0.
\end{equation}
%%%%
To satisfy the regular boundary conditions at infinity, the phase of $\omega$ is
required to satisfy 
\begin{equation}
\label{omega-phase}
 0 < \arg \sqrt{\omega^2-\mu^2}<\pi.
\end{equation}
The solution of this equation that is regular at infinity is
%%%
\begin{align}
  \label{eq:outer-sol}
  &r\psi=(-2ikr)^{l+1}e^{ikr}U(l+1-\nu+\epsilon^{2}, 2l+2+2\epsilon^{2}, -2ikr), \\
  &k=\sqrt{\omega^2-\mu^2},\quad\nu=\frac{2M\omega^{2}-M\mu^2-qQ\omega}{-ik},\notag
\end{align}
where $U$ is one of the confluent hypergeometric
functions\,\cite{abramowitz}. 
The asymptotic behavior of the solution \eqref{eq:outer-sol}
for $|kr|\ll 1$ takes the form
\begin{align}
\label{eq:outer}
  r\psi&\sim \frac{(-2ikr)^{l+1}\pi}{\sin
 \left[\pi(2l+2+2\epsilon^{2})\right]} \\
&\quad\times 
\left\{
\frac{1}{\Gamma(-l-\nu-\epsilon^{2})\Gamma(2l+2+2\epsilon^{2})}
+\cdots \right . \notag\\
&\qquad \left.
-(-2ikr)^{-2l-1}
\frac{1}{\Gamma(l-\nu+1+\epsilon^{2})\Gamma(-2l-2\epsilon^{2})}
+\cdots 
\right\}. \notag
\end{align}
In the region near the black hole horizon 
$\left(r \ll l/\mu\right)$,
Eq.\,\eqref{eq:radial} reduces to
\begin{align}
  &z\frac{d}{dz}z\frac{d}{dz}\psi+
   \left[P^2-l(l+1)\frac{z}{(1-z)^{2}}\right]\psi=0, \\
  &\quad z=\frac{r-r_+}{r-r_-},\quad
   P=-\frac{(r_{+}^2+a^2)\omega-ma-qQr_{+}}{r_{+}-r_{-}}, \notag
\end{align}
%%%
and the solution with ingoing boundary conditions at the horizon is given by
%%%
\begin{equation}
  \label{eq:inner-sol}
  \psi=z^{iP}(1-z)^{l+1}F(l+1,l+1+2iP,1+2iP,z),
\end{equation}
%%%
where $F$ is the Gauss hypergeometric function. 
The asymptotic form
of the solution \eqref{eq:inner-sol} for $1-z \ll 1$ is given by
%%%
\begin{align}
  \psi&\sim
    \frac{\Gamma(1+2iP)\Gamma(2l+1)}{\Gamma(l+1)\Gamma(l+1+2iP)}
  \left(\frac{r}{r_{+}-r_{-}}\right)^l
 +\cdots
   \notag \\
   &\quad +
    \frac{\Gamma(1+2iP)\Gamma(-2l-1)}{\Gamma(-l)\Gamma(-l+2iP)}
 \left(\frac{r}{r_{+}-r_{-}}\right)^{-l-1}
 +\cdots. 
    \label{eq:inner}
\end{align}
We get the asymptotic behavior described by \eqref{eq:outer} and
\eqref{eq:inner}. These expressions are valid in the ranges $r_{+}\ll r
\ll 1/k$ and $r_{+} \ll r \ll l/\mu$, respectively.   
Consequently, for $\omega\sim\mu$, we have the overlap region in which both the
outer expansion \eqref{eq:outer} and the inner expansion
\eqref{eq:inner} hold:
%%%
\begin{equation}
  r_{+}\ll r\ll \frac{1}{2\sqrt{\mu^2-\omega^2}}. 
\end{equation}
In this region, we can match the leading-order terms of the solutions
\eqref{eq:outer} and \eqref{eq:inner}, and this matching yields
%%%
\begin{equation}
\label{matching-condition}
\frac{\Gamma(-l-\nu-\epsilon^{2})\Gamma(2l+2)}
 {\Gamma(l-\nu+1+\epsilon^{2}) \Gamma(-2l-2\epsilon^{2})}
=
-2P\left[2k(r_{+}-r_{-})\right]^{2l+1}
\prod_{j=1}^{l}
  (j^2+4P^2)\frac{\Gamma(l+1)\Gamma(-2l-1)}{\Gamma(2l+1)\Gamma(-l)}.
\end{equation}
For $\omega \sim \mu$ and $k(r_{+}-r_{-}) \ll 1$, the  right-hand side of
the this equation is $O((k(r_{+}-r_{-}))^{2l+1})$. Therefore, we define
$\nu^{(0)}$ as the value of $\nu$ for which the right-hand side equals
zero and write $\nu\equiv \nu^{(0)}+\delta \nu$.

For $\nu=\nu^{(0)}$, the matching
condition \eqref{matching-condition} yields  
\begin{equation}
 \frac{\Gamma(-l-\nu^{(0)}-\epsilon^{2})\Gamma(2l+2)}
 {\Gamma(l-\nu^{(0)}+1+\epsilon^{2}) \Gamma(-2l-2\epsilon^{2})}=0.
\end{equation}
Thus, using the property of the gamma function $1/\Gamma(-n)=0$, we
have 
\begin{equation}
l-\nu^{(0)}+1+\epsilon^{2}=-n,
\end{equation}
where $n=0,1,2,\cdots$. From the definition of $\nu$ and
Eq.\,\eqref{eq:outer-sol}, we find
\begin{align}
 \nu^{(0)}&=
\frac{2M\omega^{(0)} \,^{2}-M\mu^{2}-qQ\omega^{(0)}}{-i\sqrt{\omega^{(0)}\,^{2}-\mu^{2}}}
\simeq l+n+1,\\ 
\omega^{(0)}&\simeq 
\mu \left[1-\left(\frac{M\mu-qQ}{l+n+1}\right)^{2}\right]^{1/2}
\simeq \mu \left[1-\frac{1}{2}\left(\frac{M\mu-qQ}{l+n+1}\right)^{2}\right],\\
 k^{(0)}&=\sqrt{\omega^{(0)}\,^{2}-\mu^{2}} 
\simeq i\frac{M\mu-qQ}{l+n+1}\mu. \label{k^{0}} 
\end{align}
From Eqs.\,\eqref{omega-phase} and \eqref{k^{0}}, to satisfy the regular
boundary conditions at infinity, we must require the condition 
\begin{equation}
  M\mu\gtrsim qQ. \label{eq:bound}
\end{equation}
With this condition, the effective potential has a well (see
Fig.\,\ref{potential}), and this condition guarantees that the
wave is confined in this well.

Next, we obtain $\delta \nu$ perturbatively. The left-hand side of
Eq.\,\eqref{matching-condition} becomes 
\begin{align}
&\frac{\Gamma(-l-\nu^{(0)}-\epsilon^{2})\Gamma(2l+2)}
 {\Gamma(l-\nu^{(0)}+1+\epsilon^{2}) \Gamma(-2l-2\epsilon^{2})}
\left[
1+\{\psi(l-\nu^{(0)}+\epsilon^{2})-\psi(-l-\nu^{(0)}-\epsilon^{2})\}\delta \nu
\right], 
\end{align}
with
\begin{align}
&\psi(z)=\Gamma'(z)/\Gamma(z). \notag
\end{align}
Then, using the property of the gamma function
\begin{equation}
 \lim_{z \rightarrow -n}\frac{\psi(z)}{\Gamma(z)}=(-1)^{n+1}n!,\quad \frac{\Gamma(-n)}{\Gamma(-m)}=\frac{(-1)^{n-m}m!}{n!},
\end{equation}
we find $\delta \nu$ as 
\begin{align}
\label{del-nu}
 \delta \nu &=2iP^{(0)}
\left[\frac{2(M\mu-qQ)}{l+n+1}\mu(r_{+}-r_{-})\right]^{2l+1} 
\times
\frac{(2l+n+1)!}{n!}\left[\frac{l!}{(2l)!(2l+1)!}\right]^{2}
\prod_{j=1}^{l}
  (j^2+4P^{(0)}\,^2), 
\end{align}
where
\begin{align}
P^{(0)}&=-\frac{(r_{+}^2+a^2)\,\omega^{(0)}-ma-qQr_{+}}{r_{+}-r_{-}}. \notag
\end{align}
Here, the relation between $\delta \nu$ and $\delta \omega$ is given by
\begin{equation}
 \delta \nu=\mu \frac{M\mu-qQ}{i}\frac{\omega^{(0)}\delta \omega}{k^{(0)}\,^{3}}
=\frac{(l+n+1)^{3}}{(M\mu-qQ)^{2}}\frac{\delta \omega}{\mu}.
\end{equation}
Thus, the real part $\sigma$ and the imaginary part
$\gamma$ of the leading-order contribution to the eigenvalue $\omega$ are 
%%%
\begin{align}
  &\sigma=\mu\left[1-\frac{1}{2}\left(\frac{M\mu-qQ}{l+1+n}\right)^2\right]
  =\mu \left[1-O(\epsilon^{2})\right], \\
  &\gamma=\mu\,\frac{\delta\nu}{i}\frac{(M\mu-qQ)^2}{(l+1+n)^3}
  =O(\epsilon^{4l+5}).
 \label{i-gamma}
\end{align}
If the imaginary part $\gamma$ of $\omega$ is positive, the
mode in question is unstable, and $\gamma$ represents the growth rate
of the scalar field.  From Eqs.\,\eqref{del-nu} and \eqref{i-gamma}, the
condition for the existence of an instability is 
%%%
\begin{equation}
 \label{eq:sr-condition}
P^{(0)}=\frac{1}{2\kappa}(m\Omega^H+q\Phi^H-\sigma)>0,
\end{equation}
%%%
where the surface gravity $\kappa$, the angular velocity $\Omega^{H}$,
and the electric potential $\Phi^{H}$ for the Kerr-Newman black hole
are introduced as follows:
%%%
\begin{equation}
 \kappa=\frac{1}{2}\left(\frac{r_{+}-r_{-}}{a^2+r_{+}^2}\right),\quad 
\Omega^{H}=\frac{a}{a^2+r_{+}^2}~,
  \quad \Phi^{H}=\frac{Qr_{+}}{a^2+r_{+}^2}~.
\end{equation}
%%%
The condition of instability \eqref{eq:sr-condition} coincides with
that of super radiance. If the super-radiant condition is
compatible with the condition for the existence of bound state \eqref{eq:bound}, the
scalar field becomes unstable. These features of the unstable mode are
consistent with the results of the analysis carried out by Damour
\textit{et al}\,\cite{damour76}. The most unstable mode corresponds to
$l=m=1$ and $n=0$, and the value of $\gamma$ for this mode is given by
%%%
\begin{equation}
  \gamma
  =\frac{\mu^4}{24}|M\mu-qQ|^5
   (a^2+r_{+}^2)^3(\Omega^{H}+q\Phi^{H}-\mu)
(\kappa^2+(\Omega^{H}+q\Phi^{H}-\mu)^2).\label{eq:gamma} 
\end{equation}
%%%
For the extreme case in which $a^2+Q^2=M^2$, $\gamma$ becomes
%%%
\begin{equation}
  \gamma=\frac{\mu^4}{24}(a^2+M^2)^3|M\mu-qQ|^5
   (\Omega^{H}+q\Phi^{H}-\mu)^3.
\end{equation}
%%%
For a Kerr black hole\,($Q=0$), $\gamma$ reduces to
%%%
\begin{equation}
  \gamma M=\frac{a}{M}\frac{(\mu M)^9}{24},
\end{equation}
%%%
which is the formula derived by Detweiler\,\cite{detweiler80}.

\begin{figure}[t]
\centering
\includegraphics[width=0.7\linewidth,clip]{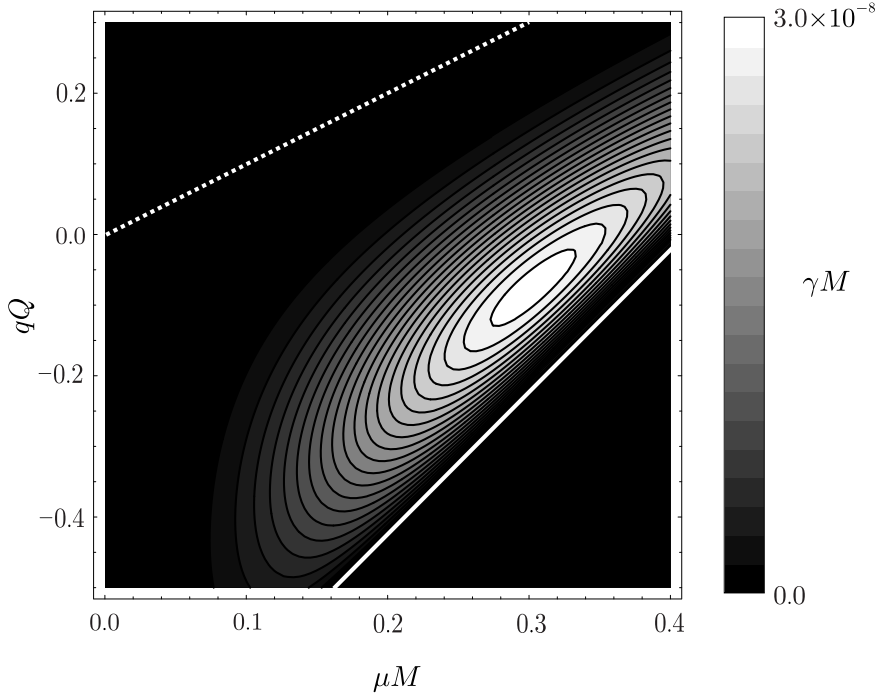}%
 \caption{The dependence of the mass $\mu$ and 
   the charge $q$ of the scalar field on the growth rate $\gamma$.
   The parameter values here are $a=0.98M,~Q=0.01M,~l=m=1$, and $n=0$. The solid line
 corresponds to 
$P^{(0)}=0$,
and the dotted line corresponds to  $M\mu-qQ=0$.
} 
\label{ana-cont}
\end{figure}

The mass and charge dependences of $\gamma$ are displayed in
Fig.\,\ref{ana-cont}. 
%%%%%%%%%%%%%%%%%%%%%%%%%%%%%%%%%%%%%%%
%%%%%%%%%%%%%%%%%%%%%%%%%%%%%%%%%%%%%%%
The maximum value is $\gamma\approx 3\times 10^{-8}$, realized at $\mu M\approx
0.3,\ qQ\approx -0.08$. The value of $\gamma$ is positive in the
region where both the super-radiant condition $P^{(0)}>0$ and the bound
state condition $M \mu \gtrsim qQ$ are satisfied. Let us consider the
region satisfying $\mu M\ll 1, |qQ|\ll 1$, where our approximation is expected to
be good. In this region, retaining only the leading-order terms of the small
parameters, $\gamma$ becomes
%%%
\begin{equation}
  \label{eq:gamma2}
  \gamma\approx \frac{\mu^4}{24}|M\mu-qQ|^5 a\,(M^2-Q^2)
\end{equation}
%%%
and the $(\mu,q)$ dependence of $\gamma$ is determined by the factor
$\mu^4|M\mu-qQ|^5$. For a fixed value of $q$, $\gamma$ is an
increasing function of $\mu$, for a fixed value of $\mu$, $\gamma$ is
a decreasing function of $qQ$. Equation \eqref{eq:gamma2} shows
that the main effect of the scalar field charge $q$ is to change the
depth of the potential well that is necessary to bound the scalar
field.  As the charge $q$ increases, the depth of the well of the scalar
field effective potential decreases, and the scalar field becomes less
bounded. Thus, an increase of the charge decreases the growth rate of
the scalar field.

%%%
For a positive charge, i.e. $q>0$, $P^{(0)}$ can be positive for a
negative azimuthal quantum number $m<0$. This contrasts with the
case of a Kerr black hole, which requires $m>0$ to realize $P^{(0)}>0$. 
This indicates the possibility of an unstable mode with $m<0$. For
$m<0$, the condition $P^{(0)}>0$ is
%%%
\begin{equation}
-|m\Omega ^{H}|M -\left(1-\frac{r_+ M}{a^2+r_+^2}\right)qQ>\mu M-qQ 
\label{eq:m<0}  
\end{equation}
%%%
for $qQ>0$.  However, the left-hand side of this inequality is negative, and
thus Eq.\,\eqref{eq:m<0} cannot be compatible with the bound state condition
\eqref{eq:bound}. Thus for $q>0$ and $m \leq 0$ in Kerr-Newman
spacetime, super radiance occurs, but the scalar field cannot be 
in a bound state, and  the mode is stable.
For Reissner-Nordstr\"om spacetime we have $a=0$, and the relation $P^{(0)}>0$ gives
%%%
\begin{equation}
 \mu <\frac{qQ}{r_{+}}.
\end{equation}
%%%
This condition also cannot be compatible with the bound state
condition \eqref{eq:bound}, and we conclude that there is no unstable
mode of the scalar field in Reissner-Nordstr\"om spacetime.
From these results, it is seen that the super radiance caused by the rotation of the
black hole is essential to make the scalar field unstable.

%%%%%%%%%%%%%%%%%%%%%%%%%%%%%%%%%%%%%%%%%%%%%%%%%%%%%%%%
\section{Numerical approach\label{sec-numerical}}
%%%%%%%%%%%%%%%%%%%%%%%%%%%%%%%%%%%%%%%%%%%%%%%%%%%%%%%%
\subsection{Method}
\begin{figure}[ht]
\centering 
  \includegraphics[width=0.6\linewidth,clip]{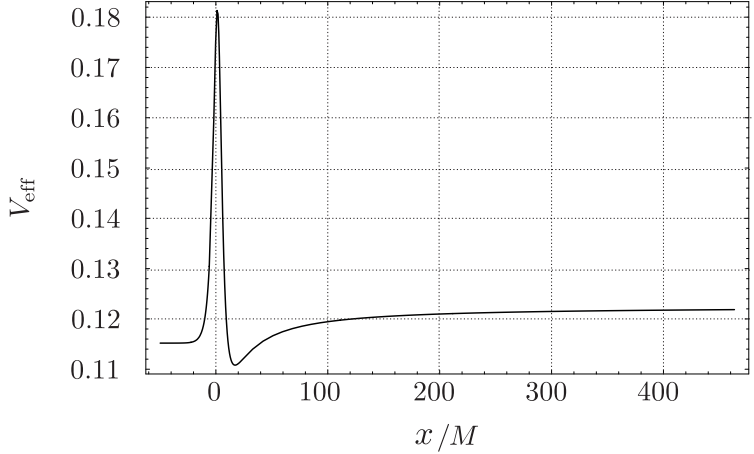}
  \caption{The effective potential $V_\text{eff}(x)$ of the scalar
    field for $a=0.98M ,~Q=0.01M,~\mu M=0.35,~qQ=-0.08,~l=m=1$.}
  \label{potential}
\end{figure}

To investigate the instability of the scalar field for a wide range
of parameter values, we carried out numerical calculations.
For this purpose, we first rewrite Eq.\,\eqref{eq:radial} in terms of the
tortoise coordinate 
%%%
\begin{equation}
  x=\int dr\frac{r^2}{\Delta}
   =r+\frac{1}{r_{+}-r_{-}}
    \left[r_{+}^2\ln(r-r_{+})-r_{-}^2\ln(r-r_{-})\right]
\end{equation}
%%%
and the new radial function $u=r\psi$. We then obtain
%%%
\begin{align}
  \label{eq:new-radial}
  \frac{d^2u}{dx^2}&=V_{\text{eff}}(r)u, \\
  \quad V_{\text{eff}}(r)&=
\frac{\Delta}{r^2}\left[\mu^2+\frac{\lambda}{r^2}+\frac{2M}{r^3}
 -\frac{2(a^2+Q^2)}{r^4}\right]
 -\frac{1}{r^4}\left[(r^2+a^2)\omega-am-qQr\right]^2, \notag \\
\quad\lambda&=l(l+1)-2ma\omega+(a\omega)^2.
 \notag
\end{align}
%%%
The effective potential $V_\text{eff}$ for $a=0.98M
,~Q=0.01M,~\mu M=0.35,~qQ=-0.08,~l=m=1$ is plotted in
Fig.\,\ref{potential}. Due to the mass of the scalar field, the
effective potential has a well, and the wave can be trapped in this
well.
%%%%%%%%% fig %%%%%%%%%%%%
%%%%%%%%%%%%%%%%%%%%%%%%%%

In the region near the horizon, the incoming solution of
Eq.\,\eqref{eq:new-radial} is given by
%%%
\begin{equation}
  u\sim \exp\left[-i\left(\frac{2M\omega-qQ}{r_{+}}-
        \frac{am+Q^2\omega}{r_{+}^2}\right)x\right], \label{eq:1}
\end{equation}
%%%
and in the far region, the regular solution of
Eq.\,\eqref{eq:new-radial} is given by
%%%
\begin{equation}
  u\sim x^{(M\mu^2-qQ\omega)/\sqrt{\mu^2-\omega^2}}
  \exp\left(-x\sqrt{\mu^2-\omega^2}\right). \label{eq:2}
\end{equation}
%%%
We use the solutions \eqref{eq:1} and \eqref{eq:2} to impose the
boundary conditions for the numerical integration of
Eq.\,\eqref{eq:new-radial}. We prepare the inner numerical boundary
$x=x_1$ near the horizon $r_{+}$ and the outer numerical boundary
$x=x_2$. By integrating Eq.\,\eqref{eq:new-radial} from $x=x_1$ with the
boundary conditions given by Eq.\,\eqref{eq:1}, we obtain a mode function
$u^{(1)}$. In the same way, we obtain $u^{(2)}$ with the boundary
conditions imposed at the far region $x=x_2$. For a given complex value
of $\omega$, if the Wronskian
%%%
\begin{equation}
  W(u^{(1)},u^{(2)})=u^{(1)}\frac{du^{(2)}}{dx}-u^{(2)}\frac{du^{(1)}}{dx}
\end{equation}
%%%
evaluated at the midpoint $x=x_m (x_1<x_m<x_2)$ is zero, the two
solutions $u^{(1)}$ and $u^{(2)}$ are linearly dependent, and $\omega$
is an eigenvalue of the equation under consideration. We search for the
zero point of the complex function $W(\omega)$ numerically in the
complex $\omega$ plane.

%%%%%%%%%%%%%%%%%%%%%%%%%%
\subsection{Result}
\begin{figure}[ht]
\centering
 \includegraphics[width=0.5\linewidth,clip]{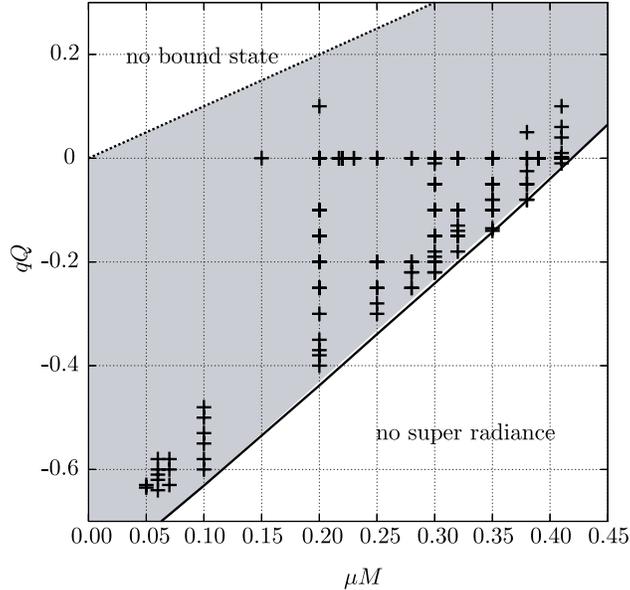}
\caption{The sets of parameter values used in the numerical calculations
 plotted corresponds to in $(\mu,q)$-space. The solid line corresponds
 to $P^{(0)}=0$ and the dotted line to $M\mu-qQ=0$. The scalar field is
 expected to be unstable for sets of parameter values in the grey region.} 
\label{dataclc}
\end{figure}

We performed numerical calculations to search for the mode of the
scalar field satisfying $l=m=1$ and $\omega \sim \mu$, for which the growth
rate of the unstable mode is expected to have the largest value. We
chose the parameter values of the black hole as $a=0.98M$ and $\ Q=0.01M$ and
used the fourth-order Runge-Kutta integrator. The numerical boundaries were set
at $x_{1}=-50M$, $x_{2}=1510M$, $x_{m}=24.5M$, and the grid spacing was
$\Delta x=0.5M$.  We obtained the value of the growth rate $\gamma$
as a function of the scalar field mass $\mu$ and the scalar field
charge $q$. We calculated the values of $\omega$ for the eigenmode using 77 different
sets of parameter values in $(\mu,q)$-space.  There points in
$(\mu, q)$-space are indicated in Fig.\ref{dataclc}.
%%%%%%%%%%%%%%%%%%%%%%%%%%%%%%%%%%%%%%
%%%%%%%%%%%%%%%%%%%%%%%%%%%%%%%%%%%%%

\begin{figure}[ht]
 \centering
  \includegraphics[width=0.6\linewidth,clip]{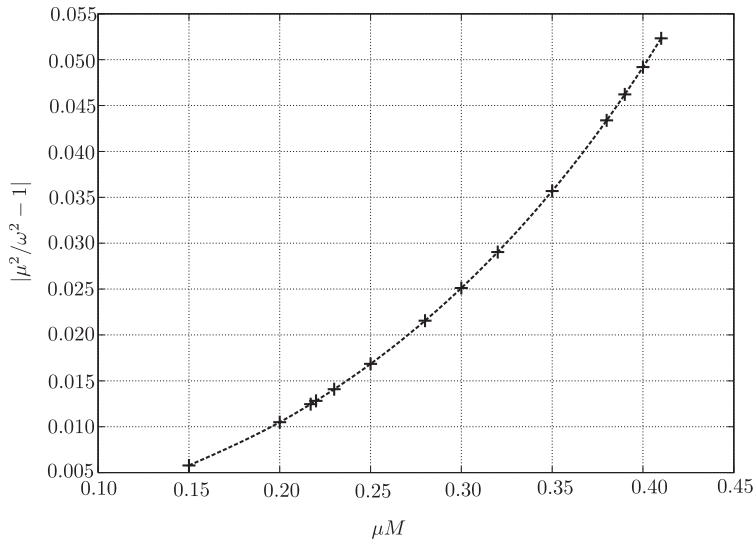}
  \caption{The value of $|\mu^2/\omega^2-1|$ as a function of $\mu$
    for $q=0$.  This value is less than $5.5\times 10^{-2}$ for all
    values of $\mu M$, and we find that our numerical results are
 consistent with the assumption $\omega \sim\mu$.} 
\label{eigen-q00-real}
\end{figure}

For the obtained value of $\omega$, we checked the validity of the assumption
$\omega\sim\mu$ used in Eq.\,\eqref{eq:new-radial}. In
Fig.\ref{eigen-q00-real}, we plot $|\mu^2/\omega^2-1|$ as a function
of $\mu$ for $q=0$. 
%%%%%%%%%%%%%%%%%%%%%%%%%%%%%%%%%%%%%%
%%%%%%%%%%%%%%%%%%%%%%%%%%%%%%%%%%%%%
For the obtained value of $\omega$, this value is smaller than
$5.5\times 10^{-2}$, and our assumption $\omega\sim\mu$ and
\eqref{eq:sepa-const} are correct for the numerically obtained modes. 
For $q\neq 0$, this value does not exceed $10^{-2}$.  We also
calculated the eigen-modes using a different grid size. To check the
numerical error on the obtained value of $\omega$, a numerical
integration with $x_2=3000M$ was also carried out. The relative error on the
value of $\omega$ evaluated from these calculations was found to be less than
$10^{-3}$.

\begin{figure}[t]
 \centering
 \includegraphics[width=0.49\linewidth,clip]{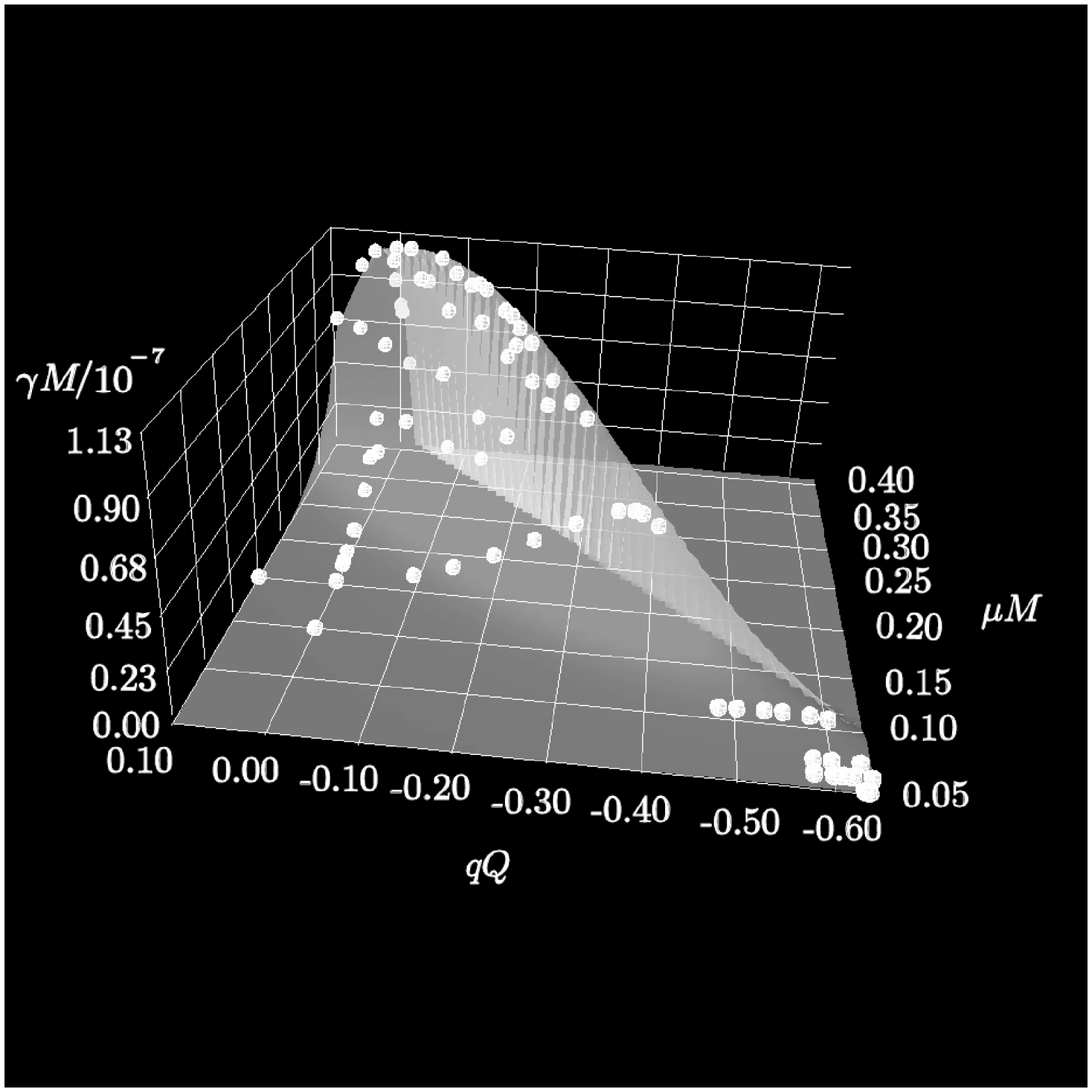}
 \includegraphics[width=0.49\linewidth,clip]{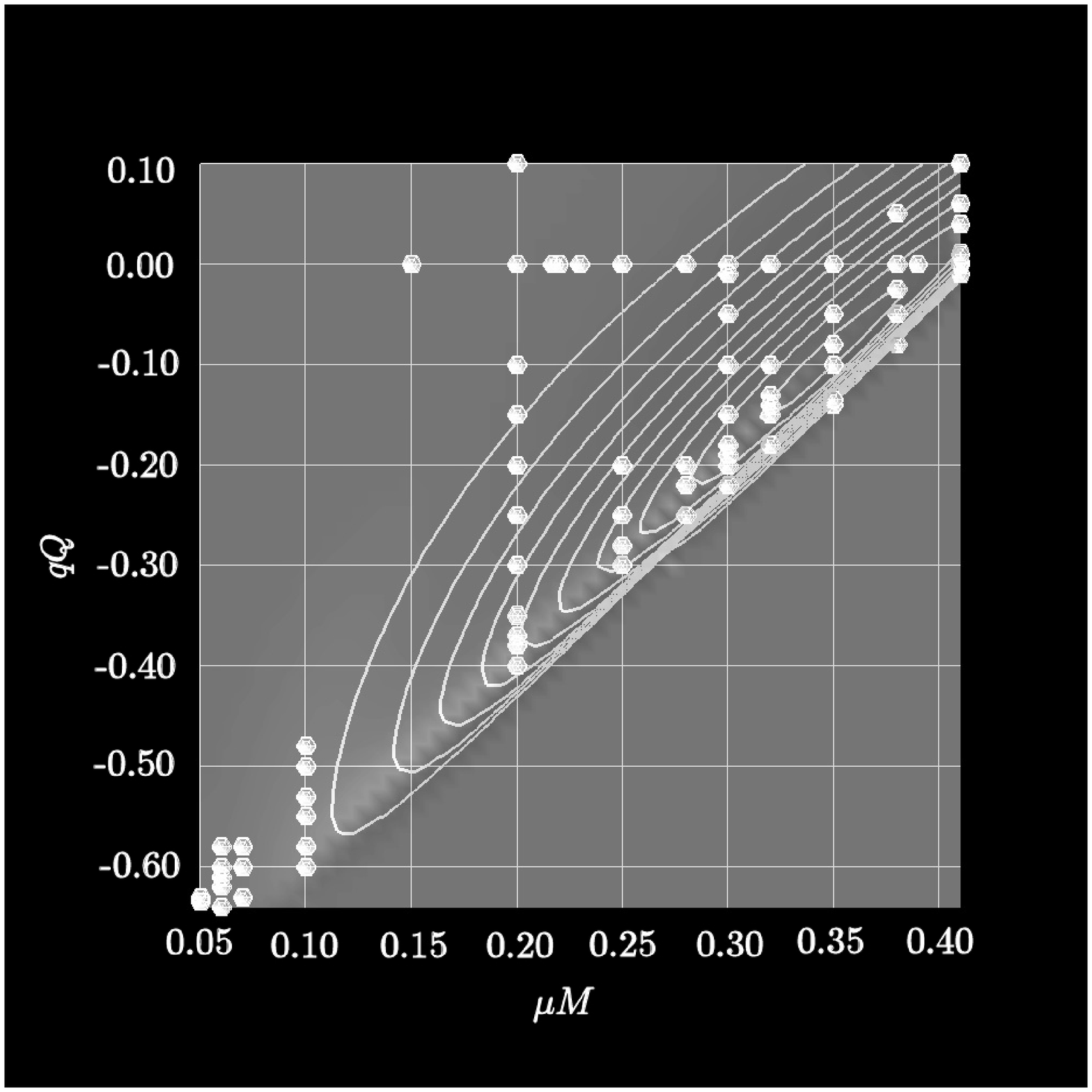}%
 \caption{The numerically obtained value of the growth rate
   $\gamma$. The spheres represent the obtained values of $\gamma$. The left
   panel is a birds-eye view of $\gamma(\mu,q)$ for
   $a=0.98M,~Q=0.01M,~l=m=1$. The right panel is the same function as
 viewed from above.}
 \label{nusuf}
\end{figure}

\begin{figure}[ht]
 \centering
  \includegraphics[width=0.496\linewidth,clip]{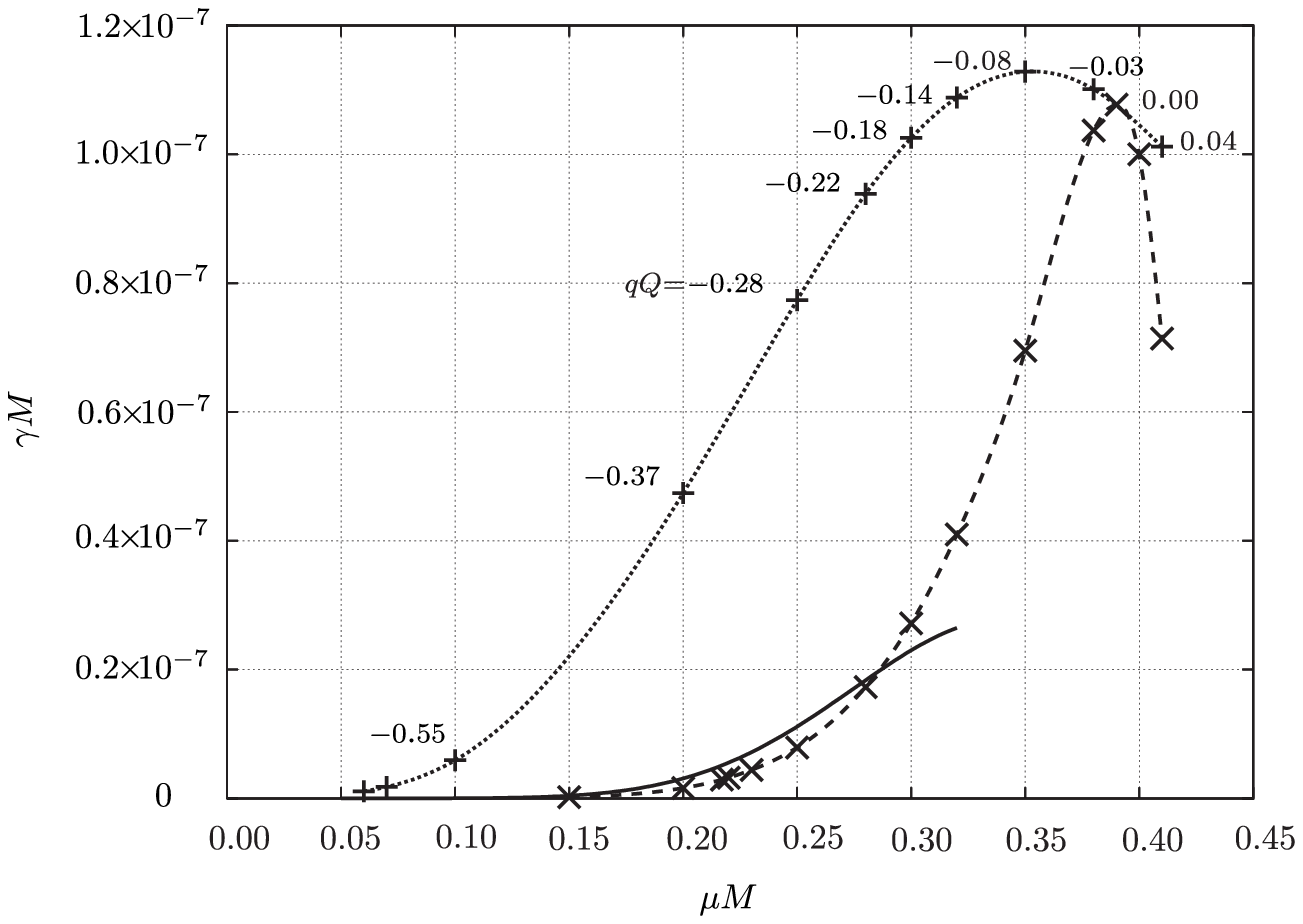}
  \includegraphics[width=0.496\linewidth,clip]{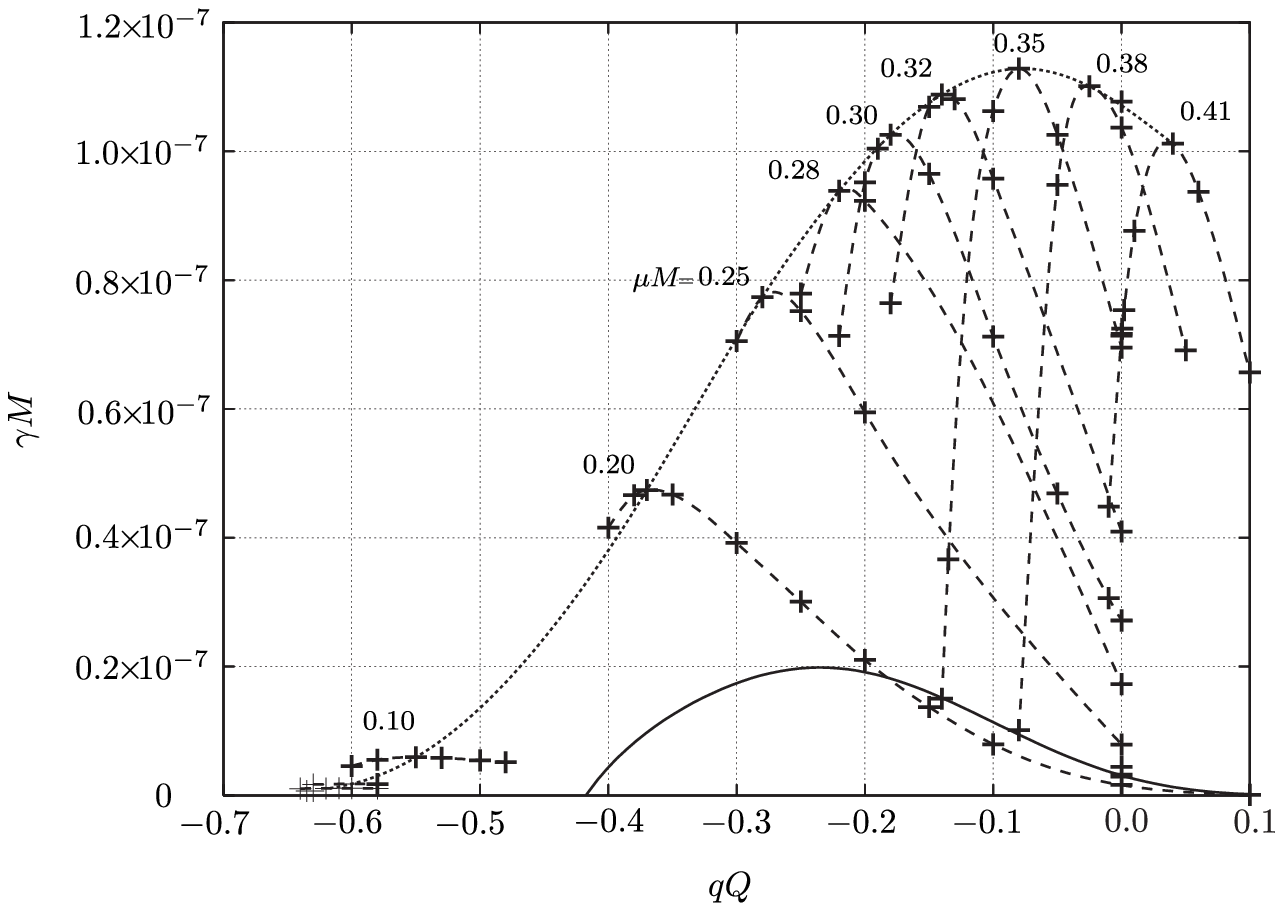} 
  \caption{The left panel displays the  $\mu$ dependence of the growth
 rate $\gamma$ for several values of $q$. The dashed curve represents the
    numerical solution for $q=0$, and the solid curve represents the analytic
    approximation \eqref{eq:gamma} for $q=0$.  The right panel plots the
 $q$ dependence of the growth rate $\gamma$ for several values of
    $\mu$.  The dashed curves represent $\gamma$ for $\mu=0.1 - 0.41$. The
    solid curve represents the analytic approximation for $\mu M=0.20$.}
  \label{2d-ana-nu}
\end{figure}

The obtained growth rate is plotted in Fig.\,\ref{nusuf}. It is seen
that the shape of the numerically obtained function $\gamma(\mu,q)$ is
almost the same as that of the analytically obtained one (see Fig.\,\ref{ana-cont}).
%%%%%%%%%%%%%%%%%%%%%%%%%%
%%%%%%%%%%%%%%%%%%%%%%%%%%%%%%%%%%%%%%%
The left panel in Fig.\,\ref{2d-ana-nu} displays the $\mu$ dependence of
$\gamma$ for different values of $q$.  The plus symbols represent the
maximum values of $\gamma$ for each $\mu$, and the cross symbols
represent $\gamma$ for $q=0$. The solid curve represents the values of
$\gamma$ for $q=0$ predicted using the analytic approximation.
%%%%%%%%%%%%%%%%%%%%%%%%%%%%%%%%%%%%%%
%%%%%%%%%%%%%%%%%%%%%%%%%%%%%%%%%%%%%
In the region satisfying $\mu M \lesssim 0.25$, the value of
the numerically obtained $\gamma$ approaches that of the analytic
solution. The deviation from the analytic solution becomes significant
for $0.3\lesssim \mu M$. For parameter values in this region, the asymptotic
matching method does not provide a good approximation.
The right panel in Fig.\,\ref{2d-ana-nu} displays the $q$ dependence of
$\gamma$. The solid curve represents $\gamma$ obtained using the analytic approximation for
$\mu M=0.2$, and it is seen that the approximation is good for $-0.2\lesssim qQ$.

The growth rate has a maximum value $\gamma M\simeq 1.13\times
10^{-7}$ at $\mu M\simeq 0.35, qQ\simeq -0.08$. The obtained minimum
value of the growth rate is $7.6\times 10^{-11}$ at $\mu M=0.20,
qQ=0.1$. Although the shape of the function $\gamma(\mu,q)$ agrees with the analytically obtained result displayed in Fig.\,\ref{ana-cont}, its
maximum value is three times larger.  We confirm that the
instability exists in the region of $(\mu, q)$-space where both the
super-radiant condition $P^{(0)}>0$ and the bound state condition
$M\mu\gtrsim qQ$ are satisfied. This parameter region is also shown in
Fig.\,\ref{dataclc}.  For all numerically obtained modes, the growth
rates are positive, and they are contained in the region bounded by the two
lines $P^{(0)}=0$ and $M\mu-qQ=0$.  As the parameter point $(\mu,q)$
approaches these lines, the growth rate decreases. We conclude that, the function
$\gamma(\mu,q)$ has a maximum value in this region. We could not
obtain a stable mode with negative $\gamma$; because the value of
$\gamma$ for the stable mode is small compared to that for the unstable
mode, it was not possible to obtain a definite value within the
accuracy of our numerical calculation.

\begin{figure}[ht]
 \centering
  \includegraphics[width=0.495\linewidth,clip]{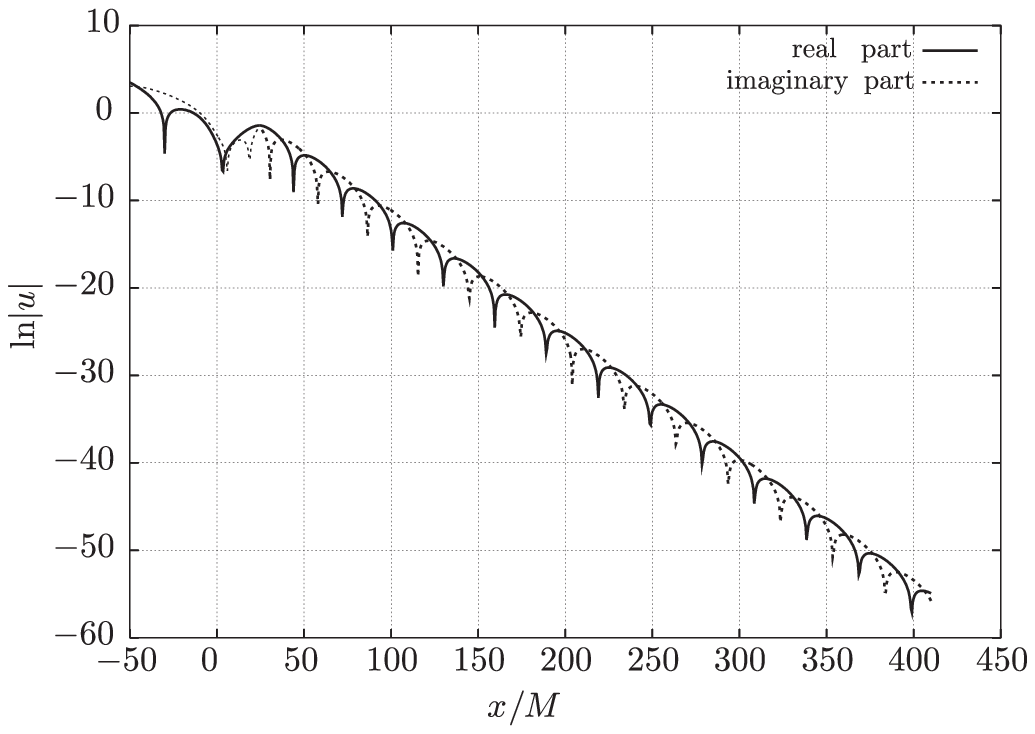}
  \includegraphics[width=0.495\linewidth,clip]{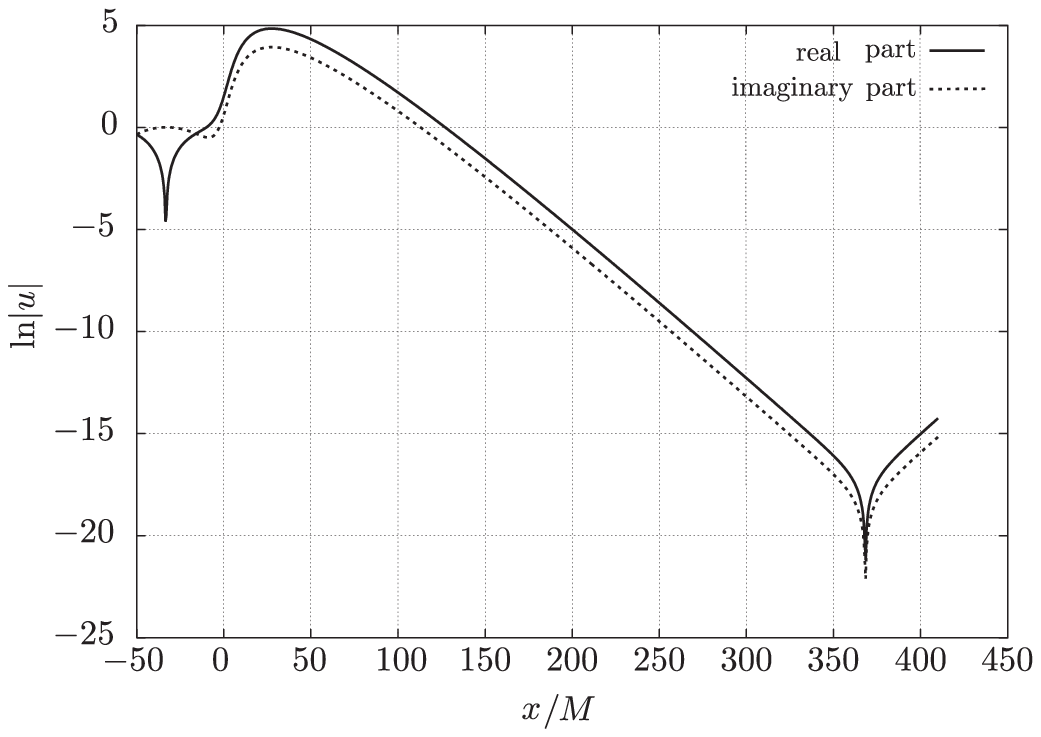}
  \caption{The behavior of the mode functions $u(x)$. The solid curve
    is the real part and the dotted curve is the imaginary part of
    $u(x)$.  The left panel plots the mode function of the stable mode
    $m=-1$. The right panel plots the mode function of the unstable mode
    $m=1$. The parameter values are $a=0.98M,~Q=0.01M,~\mu M=0.35,~qQ=-0.08,~l=1$.}
 \label{mode-function}
\end{figure}

In Fig.\,\ref{mode-function}, we show the behavior of the mode functions
for the stable and unstable cases.  
%%%%%%%%%%%%%%%%%%%%%%%%%%%%%%%%%%%%%%
%%%%%%%%%%%%%%%%%%%%%%%%%%%
The mode function of the stable mode $m=-1$ increases monotonically 
while moving toward the horizon of the black hole. Contrastingly, the mode
function of the unstable mode $m=1$ has a maximum at $x \simeq
27.5M$. This location corresponds to the minimum of the effective
potential (see Fig.\,\ref{potential}). For both values $m=\pm 1$,
the effective potential has a well, which is necessary in order for a
bound state to exist. However, for the $m=-1$ mode, the super-radiant condition is
not satisfied, and the wave falls into the black hole through the
potential barrier. For the $m=1$ mode, the super-radiant effect is
significant, and the amplified wave can accumulate in the potential
well. This leads to the instability of the mode.

%%%%%%%%%%%%%%%%%%%%%%%%%%%%%%%%%%%%%%%%%%%%%%%%%%%%%%%%
\section{Summary and discussion \label{sec-summary}}
%%%%%%%%%%%%%%%%%%%%%%%%%%%%%%%%%%%%%%%%%%%%%%%%%%%%%%%%

In this paper, we have studied the unstable modes of a massive scalar field
in Kerr-Newman spacetime. We obtained the leading-oder value of the growth
rate for $\mu M \ll1,\ qQ\ll1$ using the asymptotic matching method
and for $\mu M \lesssim 1$ using a numerical method. For a black hole with
$a=0.98M, Q=0.01M$, we obtained the maximum value of the growth rate of
the unstable mode to be $\gamma M\simeq 1.13\times 10^{-7}$ for $\mu M\simeq
0.35, qQ\simeq -0.08$. The location of the maximum value in
$(\mu,q)$-space agrees with the result of the analytic method,
but its numerical value is three times larger than that of the
analytic result. For $0.3\lesssim \mu M$, the numerically
obtained value of $\gamma$ deviates significantly from that of the analytic result
.  This indicates that the asymptotic matching method
does not give a good approximation in this parameter region.
If the mass of the scalar field is $\mu \gg 1$[MeV],
the mass of the black hole must be $M \ll 10^{15}$[Kg] for that the
instability to be maximal. Thus, there is a possibility that the
mechanism of the black hole bomb can influence the evolution of
primordial black holes.

To investigate the dynamics of the unstable mode, it is
interesting to consider the temporal evolution of a scalar wave with
an unstable mode in the black hole geometry.  The propagation of such a
wave in black hole spacetime consists of three stages. During the first
stage, there is a burst wave that depends on the initial
conditions of the wave. This stage is followed by quasi-normal
ringing and the tail mode.  If the scalar field has an unstable mode,
we expect the effect of the instability to appear in the late time
tail behavior. However, because the growth rate of the instability is
very small, it may be difficult to detect the instability
numerically. This is the next problem we intend to study.

%%%%%%%%%%%%%%%%%%%%%%%%%%%%%%%
\section*{Acknowledgements}
We would like to thank Tomoyuki Hanawa for his suggestion regarding the
numerical method used for the eigenvalue problem and Akira Tomimatsu for
valuable discussions on this subject.

%\bibliographystyle{apsrev}
%\bibliography{BH-bomb}
%
%%%%%%%%% references

\end{document}